\documentclass[oldversion]{aa}

\usepackage{graphicx}

\def\smallsun{\hbox{$_\odot$}}
\def\degr{\hbox{$^\circ$}}
\def\arcmin{\hbox{$^\prime$}}
\def\arcsec{\hbox{$^{\prime\prime}$}}
\def\cm3{cm$^{-3}$}

\begin{document}


\title{Abundances of Planetary Nebulae NGC\,3242 and
  NGC\,6369\thanks{Based on observations with the Spitzer Space
    Telescope, which is operated by the Jet Propulsion Laboratory,
    California Institute of Technology}}

\author{S.R.\,Pottasch\inst{1} \and J.\,Bernard-Salas\inst{2}}

\offprints{pottasch@astro.rug.nl} 

\institute{Kapteyn Astronomical Institute, P.O. Box 800, NL 9700 AV
  Groningen, the Netherlands \and Center for Radiophysics and Space
  Research, Cornell University, Ithaca, NY 14853}

\date{Received date /Accepted date}

\abstract{The spectra of the planetary nebulae \object{NGC\,3242 and
    NGC\,6369} are reanalysed using spectral measurements made in the
  mid-infrared with the {\em Spitzer Space Telescope} and the {\em Infrared
  Space Observatory (ISO)}. The aim is to determine the chemical composition 
  of these objects.  We also make use of {\em International Ultraviolet
  Explorer (IUE)} and ground based spectra. These elliptical PNe are 
  interesting because they are well-studied, nearby, bright objects and
  therefore allow a reasonably complete comparison of this type of nebulae.
  Abundances determined from the mid-infrared lines, which are
  insensitive to electron temperature, are used as the basis for the
  determination of the composition, which are found to differ somewhat
  from earlier results. The abundances found, especially the low value
  of helium and oxygen, indicate that the central star was originally
  of rather low mass. The abundance of phosphorus has been
  determined for the first time in NGC\,3242. The electron temperature
  in both of these nebulae is roughly constant unlike NGC\,6302 and
  NGC\,2392 where a strong temperature gradient is found.  The
  temperature of the central star is discussed for both nebulae. Finally
  a comparison of the element abundances in these nebulae with the solar
  abundance is made. The low abundance of Fe and P is noted and it is 
  suggested that these elements are an important constituent of the nebular 
  dust. }


\authorrunning{Pottasch et al.}
\titlerunning{Abundances in NGC\,3242 and NGC\,6369}  

\maketitle

\section{Introduction}

\object{NGC\,3242} (PK 261.0+32.0) is a bright planetary nebula with a
low radial velocity and is located at a rather high galactic latitude.
A photograph of the nebula is shown in Fig.\,1. The nebula has a
bright inner ellipsoidal shell of about 28\arcsec x 20\arcsec. This is
surrounded by an almost spherical halo of 46\arcsec x 40\arcsec with
considerably lower emission. Both the inner and outer regions contain
little structure.  The nebula is located about 32 degrees above the
galactic plane and has little or no extinction. Because of its
brightness it is clear that it is a nearby nebula. An expansion
distance is known for this nebula to be about 0.5 kpc (Terzian,
\cite{terz}; Mellema, \cite{mell}).

The nebula has a rather bright central star (V=12.43) which has been
studied by several authors. Pauldrach et al. (\cite{paul}) have
compared the stellar spectrum with a model atmosphere and conclude
that the star has an effective temperature T$_{eff}$=75\,000 K which
is in between the hydrogen Zanstra temperature of about 59\,000 K, and
somewhat less than the ionized helium Zanstra temperature which is
close to 91\,000 K (Gathier \& Pottasch, \cite{gath}).  Tinkler \&
Lamers (\cite{tinkler}) prefer to assign a higher effective
temperature of T$_{eff}$=94\,000 K. to the star based on the high
ionized helium Zanstra temperature. We shall discuss the stellar
temperature later in the paper.

NGC\,6369 (PK 002.4+05.8) is located in the direction of the galactic
bulge but it is undoubtably a nearby PN for two reasons. First, it is
a rather large nebula, with a diameter of about 32\arcsec. Second, it
is the third brightest PN in the sky, at least at radio frequencies.
It cannot however be very close because the extinction is quite high
and its radial velocity is also high (-101 km/s). The nebula can be
described as a bright ring with an outer diameter of 32.4\arcsec and
an inner diameter of about 13\arcsec. There is faint emission both to
the east and the west of the main structure. The nebula is shown in
Fig.\,2. The central star is clearly visible. It is classified as
spectral type WC4 and has a magnitude V=15.91 (Gathier \&
Pottasch \cite{gath}). Its hydrogen Zanstra temperature is about
70\,000 K and ionized helium Zanstra temperature is 106\,000 K
(Monteiro et al. \cite{mont}). Judging from these temperatures it
would appear that the central star of NGC6369 is somewhat hotter than
that of NGC3242. Yet as we shall see, the HeII lines are considerably
stronger in NGC3242. This will be discussed further in Sect.6 on the
basis of the excitation of other elements found in the nebulae.

The purpose of this paper is to study the element abundances in these
nebula with the help of mid-infrared spectra, in the hope that the
chemical abundances will shed some light on the evolution of these
nebulae. In recent years abundances in NGC\,3242 have been studied by
Tsamis et al. (\cite{tsamis}), by Barker (\cite{barker}), by Henry et
al.  (\cite{henry}), by Aller \& Czyzak (\cite{aller}) and by Krabbe
\& Copetti (\cite{krab}). For NGC\,6369 abundances have been studied
by Aller \& Keyes (\cite{ allerkey}), Pena et al. (\cite{pena}) and
Monteiro et al. (\cite{mont}). All of these groups use optical
nebular spectra (taken by themselves) and in the case of NGC\,3242
ultraviolet {\em IUE} spectra were also used. For NGC\,6369
ultraviolet {\em IUE} spectra were taken but because of the very high
extinction in the direction of this nebula they are underexposed and
unusable.  Barker (\cite{barker}) has measurements of NGC\,3242 taken
at five different positions in the nebula. He uses low dispersion {\em
  IUE} measurements taken with the small aperture (3\arcsec~diameter)
which are very noisy.  Henry et al. (\cite{henry}) use low dispersion
{\em IUE} measurements taken with the large aperture (10\arcsec x
23\arcsec).

We have measured the spectrum of both NGC\,3242 and NGC\,6369 in the
mid-infrared with the IRS spectrograph of the {\em Spitzer Space
  Telescope} (Werner et al. \cite{werner}) and with the ISO
spectrograph. The use of the mid-infrared spectrum permits a more
accurate determination of the abundances.  The reasons for this have
been discussed in earlier studies (e.g. see Pottasch \& Beintema
\cite{pott1}; Pottasch et al.  \cite{pott2}, \cite{pott4}; Bernard
Salas et al. \cite{bernard}), and can be summarized as follows.

1) The intensity of the infrared lines is not very sensitive to the electron
temperature nor to possible extinction effects.
2) Use of the infrared line intensities enable a more accurate determination
of the electron temperature for use with the visual and ultraviolet lines.
3) The number of observed ionization stages is doubled.

This paper is structured as follows. First the Spitzer and {\em ISO}
spectra of both nebulae are presented and discussed (in Sect.\,2).
Then the intrinsic H$\beta$ flux is determined using both the
measurements of the infrared hydrogen lines and the radio continuum
flux density (Sect.\,3). The visible spectrum of the nebula is
presented in Sect. 4 together with a new reduction of the ultraviolet
({\em IUE}) spectrum of NGC\,3242 This is followed by a discussion of
the nebular electron temperature and density and the chemical
composition of both nebulae (Sect.\,4). A comparison of the resulting
abundances with those in the literature is given in Sect.\,5. In
Sect.\,6 a possible evolution of these nebulae is compared to other
PNe. In Sect.\,7 the central star is discussed especially in relation
to the nebular spectrum.  In Sect.\,8 a general discussion and
concluding remarks are given.

\section{The infrared spectrum}

Observations of both NGC\,3242 and NGC\,6369 were made using the
Infrared Spectrograph (IRS, Houck et al. \cite{houck}) on board the
{\em Spitzer Space Telescope} with AORkeys of 16463360 and 4905216
respectively. NGC\,6369 was observed in staring mode using the SH
module (9.5-19.5~$\mu$m, R$\sim$600). In this observing mode the
spectrum is taken in two positions at 1/3 and 2/3 of the length of the
slit which are referred as the nod positions. The reduction in this
case started from the {\em droop} images which are equivalent to the
most commonly used Basic Calibrated Data ({\em bcd}) images but lack
stray-cross removal and flatfield.  NGC\,3242 was observed in cluster
mode at 3 positions, one centered at the target and two for background
of which only the closer sky position relative the target was used,
with the SL, SH, and LH modules (5.4-37~$\mu$m). In cluster mode the
observations are taken in the center of the slit, as opposed to the
nod positions. Because the stars used for calibration were observed at
the nod positions in this case the reduction started from the {\em
  bcd} images.  The data were processed using the s15.3 version of the
pipeline and using a script version of {\em Smart} (Higdon et al.
\cite{higdon}).  The tool {\em irsclean} was used to remove rogue
pixels. The different cycles for a given module were combined to
enhance the S/N.  At this point the background images were subtracted
to remove the sky contribution in NGC\,3242. We note that since we are
interested in line fluxes the removal of background is irrelevant for
our analysis except for aiding in the removal of any rogue pixel that
may have been left out by the {\em irsclean} tool (i.e. those with low
flux). Then the resulting HR images were extracted using full aperture
measurements and the SL module in NGC3242 using a fix column
extraction (10 pixels). 

The great advantage of the IRS spectra compared to the ISO SWS spectra
is the very high sensitivity of the IRS.  Otherwise the two
instruments are comparable. The IRS high resolution spectra have a
spectral resolution of about 600, which is a factor of between 2 and 5
less than the resolution of the ISO SWS spectra.

  \subsection{Diaphragm sizes}

  The mid-infrared measurements are made with several different
  diaphragm sizes.  Because most of the diaphragms are smaller than
  the size of the nebulae we are presently studying, we first discuss
  how the different spectra are placed on a common scale.

  The spectra made with the IRS are taken with three different
  diaphragms: the IRS high resolution instrument (spectral resolution
  of about 600) measures in two spectral ranges with two different
  modules: the short high module (SH) measures from 9.9$\mu$m to
  19.6$\mu$m and the long high module (LH) from 18.7$\mu$m to
  37.2$\mu$m. The SH has a diaphragm size of 4.7\arcsec x 11.3\arcsec,
  while the LH is 11.1\arcsec x 22.3\arcsec. If the nebulae are
  uniformly illuminating then the ratio of the intensities would
  simply be the ratio of the areas measured by the two diaphragms.
  Since this is probably not so (because of low intensity holes in the
  nebulae), we may use the ratio of the continuum intensity in the
  region of wavelength overlap at 19$\mu$m. There are other ways of
  determining this ratio which will be discussed presently. The SL
  (low resolution) spectra are made with a long slit which is 4\arcsec
  wide and extends over the entire nebula.

  The ISO diaphragms are somewhat larger. The SWS measurements below
  12$\mu$m are made with a diaphragm 14\arcsec x 20\arcsec; between
  12$\mu$m and 27$\mu$m it is somewhat larger (14\arcsec x 27\arcsec)
  and above 27$\mu$m it is 20\arcsec x 33\arcsec. The ISO LWS spectra
  (which cover a spectral region from 45$\mu$m to almost 200$\mu$m)
  are taken with a diaphragm which has a diameter of about 80\arcsec
  and covers the entire nebula.

  \subsubsection{NGC\,3242}

  \begin{figure}
    \centering
      \includegraphics[width=8cm,angle=0]{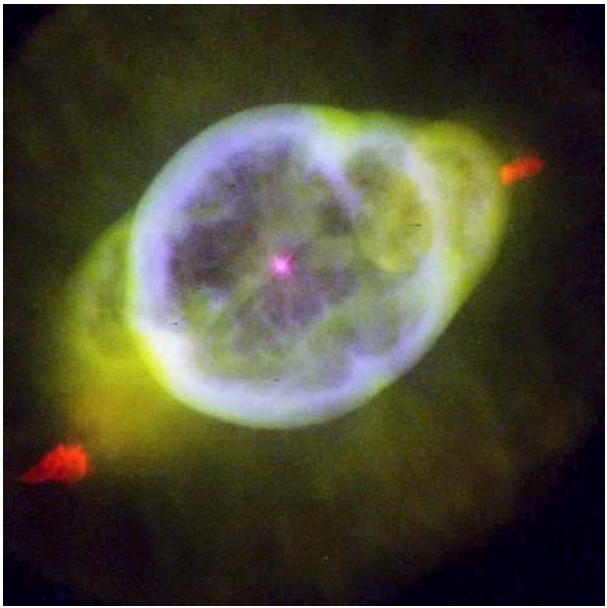}
      \caption{An HST image of NGC\,3242. Credit B.Balick et al.}
    \label{m142_f}
  \end{figure}


  The IRS measurement of NGC\,3242 was centered at RA(2000)
  10$^{h}$24$^{m}$46.11$^{s}$ and Dec(2000)
  -18\degr38\arcmin32.6\arcsec. This is almost the same as the value
  measured by Kerber et al. (\cite{kerber}) of RA(2000)
  10$^{h}$24$^{m}$46.138$^{s}$ and Dec(2000)
  -18\degr38\arcmin32.26\arcsec, which is presumably the coordinate of
  the central star. Thus the IRS measurement was well centered on the
  nebula and both the LH and SH diaphragms measured the inner,
  brighter nebula. Since the LH diaphragm is larger, more of the
  nebula is seen in this diaphragm and therefore a correction must be
  made to bring the two measurements to the same scale. This was done
  first by making use of the fact that the two spectrographs had a
  small wavelength region in common at about 19$\mu$m. To make the
  continuum emission at this wavelength equal, the SH emission had to
  be increased by 3.96. The ratio can also be determined by using the
  ratio of lines of a given ion which are observed both with the SH
  and LH diaphragms and at the same time are also insensitive to the
  electron temperature and density. Such a pair of lines is from
  \ion{[Cl}{iv]} at 11.76 and 20.31$\mu$m, which give about the same
  factor between the SH and LH intensities. Normally the
  \ion{[Ne}{iii]} lines could also be used to obtain this ratio but
  this was not used in this case because the intensity of the
  \ion{[Ne}{iii]} line at 36.01$\mu$m is badly determined.

  The measured emission line intensities for NGC\,3242 are given in
  Table 1, after correcting the SH measurements by the factor 3.96 in
  the column labeled 'intensity'.  The fluxes were measured using the
  Gaussian line-fitting routine.
  The last column gives the ratio of the intensity to H$\beta$ where
  the H$\beta$ is found from the strongest hydrogen line(s) measured
  in the IRS spectrum. It has been assumed that the ratio of the sum
  of the two hydrogen lines (n=7-6 and n=11-8) at 12.372$\mu$m to
  H$\beta$ has a value of 10.15 x 10$^{-3}$, which is given by Hummer
  \& Storey (\cite{hummer}) for an electron temperature of 12\,500 K.
  No correction for extinction is made since it is very small. Other
  hydrogen lines listed in the table can be used as well but the line
  at 11.306$\mu$m is blended with the \ion{He}{ii} transition
  n=18-14. All the lines predicted a value of H$\beta$ through the SH
  diaphragm of 2.90x10$^{-11}$ erg cm$^{-2}$s$^{-1}$ which translates
  to a value of 1.148x10$^{-10}$ ergs cm$^{-2}$s$^{-1}$ in the LH
  diaphragm. The \ion{[Ar}{iii]} line at 8.992$\mu$m was measured with
  the low resolution spectrograph (SL) using a diaphragm of 18\arcsec
  x 3.6\arcsec. The intensity has been normalized so that the other
  four lines measured in SL have the same intensities as these lines
  in SH.

\begin{table}
  \caption[]{IRS spectrum of \object{NGC\,3242}. The measured line intensity is
    given in Col.3. The last column gives the ratio of the line
    intensity to H$\beta$(=100).}

\begin{center}
\begin{tabular}{lrcc}
\hline
\hline
Identification &  $\lambda$($\mu$m) & Intensity$^{\dagger}$ & I/H$\beta$ \\
\hline

\ion{[Ar}{iii]} & 8.992 & 904$\pm$30 & 8.48 \\
\ion{[S}{iv]} & 10.511 & 8220$\pm$160  & 70.3\\
\ion{H}{i}(9-7) & 11.306 & 38.4$\pm$2.3 & 0.22 \\
\ion{[Cl}{iv]} & 11.760 & 70.8$\pm$3.5 & 0.616 \\
\ion{H}{i} (7-6+11-8) & 12.370 & 112.1$\pm$4.6 &    \\
\ion{[Ne}{ii]} & 12.810 & 150$\pm$3.9  & 1.31 \\
\ion{[Ar}{v]} & 13.099 & 81.5 $\pm$4.7 &  0.710   \\
\ion{[Ne}{v]} & 14.319 & 35.6$\pm$1.6 & 0.31 \\
\ion{[Ne}{iii]} & 15.553 & 13320$\pm$216 & 116.0 \\
\ion{H}{i} (10-8) & 16.203 & 14.2$\pm$3.4 & 0.15 \\
H$_{2}$    &  16.881 & 17.3$\pm$9.7 &     \\
H$_{2}$    &  16.976 & 5.2$\pm$3.3 &       \\
\ion{[P}{iii]}  & 17.890 & 37.0$\pm$2.0 & 0.322 \\
\ion{[S}{iii]} & 18.711 & 957$\pm$22  & 8.33 \\
\ion{H}{i} (8-7) & 19.065 & 43.0$\pm$5.0 & 0.401 \\
\ion{[Cl}{iv]} & 20.312 & 57.0$\pm$2.66  & 0.497 \\
\ion{[Ar}{iii]} & 21.818 & 39.6$\pm$6.4 & 0.345 \\
\ion{[O}{iv]} & 25.884 & 15724$\pm$278  & 137 \\
\ion{[S}{iii]} & 33.473 & 363:$\pm$22  & 3.16: \\
\ion{[Ne}{iii]} & 36.014 & 785:$\pm$98 & 6.84: \\ 

\hline
\ion{[O}{iii]}  & 51.8 & 34700$\pm$1400 & 162 \\
\ion{[N}{iii]}  & 57.3 & 3100$\pm$140   & 14.5 \\
\ion{[O}{iii]}  & 88.4 & 15500$\pm$430  & 72.4 \\
\hline

\end{tabular}
\end{center}

$^{\dagger}$ Intensities measured in units of 10$^{-14}$
erg~cm$^{-2}$~s$^{-1}$.  The intensities below 19$\mu$m have been
increased by a factor of 3.96 to bring them on the same scale as the LH
intensities measured through a larger diaphragm. The intensities below 10$\mu$m
are measured with the low resolution instrument.
 
: Indicates an uncertain value.
 
The last three measurements are taken from ISO.

\end{table}

  There are no ISO SWS measurements of NGC\,3242 because it was not
  often visible to the satellite. There is however one ISO LWS
  measurement and the resulting line intensities are listed in the last
  three lines of Table 1 (taken from Liu et al. (\cite{liulws}).
  Because the LWS had a diaphragm of almost 80\arcsec, the intensities
  given are for the entire nebula. To obtain the ratio of the
  intensities to H$\beta$ (the last column) the intensities were divided
  by the value of H$\beta$ obtained from the 6cm radio continuum
  emission given in the next section.

\subsubsection{NGC\,6369}

The IRS measurement of NGC\,6369 was centered at RA(2000)
17$^{h}$29$^{m}$20.78$^{s}$ and Dec(2000)
-23\degr45\arcmin32.3\arcsec. This is almost the same as the value
measured by Kerber et al. (\cite{kerber}) of RA(2000)
17$^{h}$29$^{m}$20.443$^{s}$ and Dec(2000)
-23\degr45\arcmin34.2\arcsec, which again is presumably the coordinate
of the central star. The IRS measurement was again well centered on
the nebula and the SH diaphragm measured the inner, brighter nebula
and part of the less bright hole. We do not have measurements made with
the LH diaphragm. We do however have ISO measurements of this nebula.
They were centered at almost exactly the same position as the IRS
measurements. The IRS measurements are listed in Table 2 which is
arranged in a similar way as Table 1. The intensities listed in
Col.\,3 are those measured with the SH diaphragm. The ratio of the
intensity to H$\beta$ shown in the last column is found in the
following way. The H$\beta$ used is derived from the hydrogen lines
listed in the table which are measured through the same diaphragm.
They are interpreted in terms of H$\beta$ using the theoretical ratios
given by Hummer \& Storey (\cite{hummer}) for an electron temperature
of 12\,500 K (see Sect.\,4). Furthermore because the extinction is
high in this nebula a correction for this is also made in this column.
The correction is small in the infrared.

  \begin{figure}
    \centering
      \includegraphics[width=8cm,angle=0]{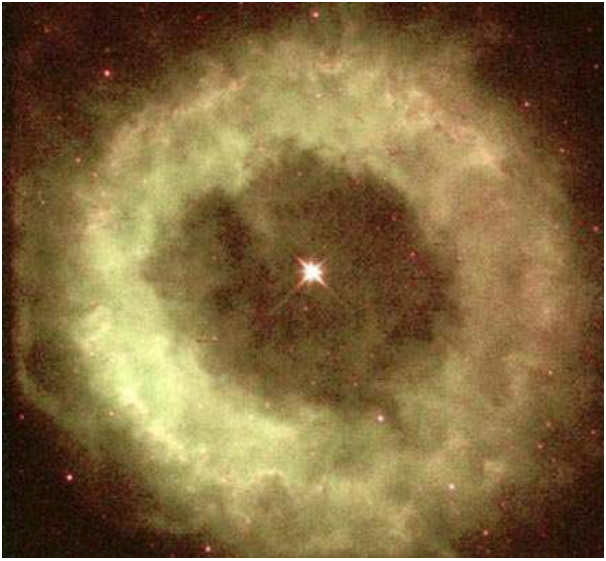}
      \caption{An HST image of NGC\,6369. Credit H.Bond \& R.Ciardullo.}
    \label{m142_f}
  \end{figure}

\begin{table}
  \caption[]{IRS spectrum of \object{NGC\,6369}. The measured line intensity is
    given in Col.3. The last column gives the ratio of the line
    intensity to H$\beta$(=100).}

\begin{center}
\begin{tabular}{lrcc}
\hline
\hline
Identification &  $\lambda$($\mu$m) & Intensity$^{\dagger}$ & I/H$\beta$ \\
\hline

\ion{[S}{iv]} & 10.505 & 2110$\pm$66  & 65.2\\
\ion{H}{i}(9-7) & 11.301 & 14.2$\pm$0.6 &  \\
\ion{[Cl}{iv]} & 11.755 & 9.95$\pm$0.7 & 0.308 \\
\ion{H}{i} (7-6+11-8) & 12.365 & 31.3$\pm$1.3 &    \\
\ion{[Ne}{ii]} & 12.806 & 147$\pm$6.0  & 4.14 \\
\ion{[Ar}{v]} & 13.10 & $\leq$0.90  & $\leq$ 0.026   \\
\ion{[Mg}{v]} & 13.52 & $\leq$1.27 & $\leq$0.035 \\
\ion{[Cl}{ii]} & 14.356 & 2.13$\pm$0.41   & 0.060 \\
\ion{[Ne}{iii]} & 15.546 & 4971$\pm$147 & 139.0 \\
\ion{H}{i} (10-8) & 16.198 & 4.29$\pm$0.24 &    \\
H$_{2}$    &  17.036 & 2.63$\pm$0.15 &       \\
\ion{[P}{iii]}  & 17.882 & 28.5$\pm$0.9 & 0.814 \\
\ion{[S}{iii]} & 18.703 & 870$\pm$24  & 24.8 \\
\ion{H}{i} (8-7) & 19.056 & 11.1$\pm$0.27 &   \\
\hline

\end{tabular}
\end{center}

$^{\dagger}$ Intensities measured in units of 10$^{-14}$
erg~cm$^{-2}$~s$^{-1}$ and are as measured through the SH diaphragm
 
: Indicates an uncertain value.
 
  \end{table}

  The ISO measurements of NGC\,6369 are shown in Table 3. As can be
  seen from the table, the wavelength covered is considerably large
  than the IRS measurements but because the sensitivity is lower only
  the stronger lines are seen. The intensities measured are given in
  Col.3 of the table and are accurate to about 20\%.  Because the
  extinction, which is given in the next section, is large, the
  intensities corrected for extinction are  given in Col.4. In the last
  column the ratio of the corrected intensity to the value of H$\beta$
  which would have been measured through the same diaphragm (and
  corrected for extinction). A few words must be said about the values
  of H$\beta$ used because it varies with the diaphragm used. The
  value of H$\beta$ used below 12$\mu$m is found from the Brackett
  $\alpha$ and $\beta$ lines. Using an electron temperature of
  T=12\,500 K this predicts a value of H$\beta$=1.53 x 10$^{-10}$
  erg~cm$^{-2}$~s$^{-1}$ in this region. For the spectral regions
  above 12$\mu$m we have increased the H$\beta$ flux by the ratio of
  the diaphragm size; thus from 12$\mu$m to 27$\mu$m we find 2.06 x
  10$^{-10}$ erg~cm$^{-2}$~s$^{-1}$ and above this wavelength (and
  less than 40$\mu$m) we use 3.60 x 10$^{-10}$ erg~cm$^{-2}$~s$^{-1}$.
  This is the correct procedure as long as the nebula is larger than
  the largest diaphragm used and the emission is uniform. This may be
  checked by comparing the continuum emission ratios measured at the
  transition wavelengths. This gives a consistent result although the
  continuum is rather noisy, especially at 12$\mu$m. Another manner of
  check is to note that the ratio of the \ion{[Ne}{iii]} lines
  15.5/36.0, which has only a very small dependence on electron
  temperature and density, has its predicted theoretical value. For
  the wavelength region above 50$\mu$m the diaphragm is large enough
  to include the entire nebula; therefore The value of H$\beta$ used
  is that found from the 6\,cm flux density: 6.16 x 10$^{-10}$
  erg~cm$^{-2}$~s$^{-1}$ (see the following section).

  The values of I/H$\beta$ found from the IRS and the ISO measurements
  can be compared for four lines in common to the two instruments (see
  Tables 2 and 3).  In three of the four cases agreement is within
  10\%. Only for the \ion{[S}{iv]} line is the agreement less good. We
  suggest that the difficulty lies with the IRS measurement because
  this part of the spectrum is badly calibrated. The line measured in
  different orders differs by about 34\%. In the discussion below the
  ISO value will be used for this line and the IRS values for the
  other lines.

\begin{table}
  \caption[]{ISO spectrum of \object{NGC\,6369}. The measured line intensity is
    given in Col.3 and is corrected for extinction in Col.4. The last column 
    gives the ratio of the line intensity to H$\beta$(=100).}

\begin{center}
\begin{tabular}{lrccc}
\hline
\hline
Identification &  $\lambda$($\mu$m) & Intensity$^{\dagger}$ & Corr. Intens. &
I/H$\beta$ \\
\hline

\ion{H}{i}(6-4) & 2.625  & 519   & 649  &     \\
?               & 2.846  & 341  & 426  & 2.07  \\
\ion{H}{i}(5.4) & 4.051  &  978  & 1086 &      \\
\ion{[Ar}{iii]} & 8.988 & 1940 & 2390 & 15.6 \\
\ion{[S}{iv]} & 10.507 & 11570  & 13800 & 90.2 \\
\ion{[Ne}{ii]} & 12.806 & 763 & 835 & 4.05 \\
\ion{[Ne}{iii]} & 15.549 & 27800 & 30000 & 145.6 \\
\ion{[S}{iii]} & 18.705 & 5350  & 5890 & 28.6 \\
\ion{[O}{iv]} & 25.884 & 2045  & 2390 & 10.6 \\
\ion{[S}{iii]} & 33.469 & 4060  & 4220 & 11.7 \\
\ion{[Ne}{iii]} & 35.992 & 4820 & 4980 & 13.8 \\ 
\ion{[O}{iii]}  & 51.822 & 68600 &     & 111.4 \\
\ion{[N}{iii]}  & 57.306 & 10400 &     & 16.9 \\
\ion{[O}{i]}    & 63.163 & 5850  &     &  9.5 \\
\ion{[O}{iii]}  & 88.351 & 21360  &    & 34.7 \\
\ion{[N}{ii]}   & 121.905 & 201  &     &  0.33 \\
\ion{[O}{i]}    & 145.486 & 155  &     &  0.25 \\
\ion{[C}{ii]}   & 157.712 & 674  &     &  1.09 \\
\hline

\end{tabular}
\end{center}

$^{\dagger}$ Intensities measured in units of 10$^{-14}$
erg~cm$^{-2}$~s$^{-1}$ and have been corrected for extinction in Col.4.  The 
determination of I/H$\beta$ is discussed in the text.
 
: Indicates an uncertain value.
 
\end{table}

\section{Extinction}

The extinction may be found both by a comparison of radio emission with 
H$\beta$ flux and by a comparison of observed and
theoretical Balmer decrement, both of which we will discuss.

  \subsection{The 6\,cm radio emission}

  The 6\,cm flux density for both nebulae has been measured by
  Griffith et al.  (\cite{grif}) in the Parkes-MIT-NRAO survey. Their
  reduction gave two values: one using a Fixed width fit and the other
  a General width fit. For NGC\,6369 they find 2041 and 1893 mJy and
  for NGC\,3242 760 and 731 mJy. In addition NGC\,6369 was measured by
  Milne \& Aller (\cite{milne}) who find a value of 2002 mJy for this
  object. We will use a value of 2000 mJy for NGC\,6369 which, using
  values of T$_e$ and helium abundance determined below together with
  the equation quoted in Pottasch (\cite{potta}), implies a total
  H$\beta$ flux of 6.16x10$^{-10}$ erg~cm$^{-2}$~s$^{-1}$ for this
  object.  Similarly a value of 745 mJy for NGC\,3242 implies a total
  H$\beta$ flux of 2.14x10$^{-10}$ erg~cm$^{-2}$~s$^{-1}$ for this
  object.

  \subsection{Extinction}

  The measured value of the integrated H$\beta$ flux for NGC\,6369 and
  NGC\,3242 is 4.76x10$^{-12}$ and 1.62x10$^{-11}$
  erg~cm$^{-2}$~s$^{-1}$ respectively (see Acker et al.,
  \cite{acker}). Using the value of H$\beta$ from the radio
  measurements leads to a value of C=2.116 or
  $E_{\mathrm{B-V}}$=1.444. For NGC\,3242 these values are C=0.12
  $E_{\mathrm{B-V}}$=0.083.

  The extinction determined from the Balmer decrement differs somewhat
  according to different authors. For NGC\,6369 Monteiro et
  al. (\cite{mont}) find C=2.17, Aller \& Keyes (\cite{allerkey})
  give C=2.23 and Pena et al.  (\cite{pena}) find C=1.9. For NGC\,3242
  Henry et al. (\cite{henry}) give C=0, Aller \& Czyzak (\cite{aller})
  give C=0.14, and Barker (\cite{barker}) finds C=0.15. These values
  are in rough agreement with those found from the radio/H$\beta$
  method. Since this method is the most accurate determination, this
  value will be used when necessary in this paper.

  \subsection{The visual spectrum}

  The visual spectrum has been measured by several authors for each of
  the nebulae. We list here the results of three of the most recent
  high resolution spectra for each nebula. The line intensities
  reported have been corrected by each author for a value of
  extinction determined by them to obtain a theoretically correct
  Balmer decrement. The result are listed in Tables 4 and 5, where the
  last column lists the average value which we have used. No attempt
  has been made to use a common extinction correction because then the
  Balmer decrement will be incorrect. The value of extinction C which
  the individual authors found is listed at the bottom of the table.
  All authors estimate that the strongest lines have a 10\% error, the
  intermediate strength lines (about 5\% of H$\beta$) have about 20\%
  error and the weakest lines have about 30\% error. The average
  intensities have about the same error.

  The measured optical line fluxes of NGC\,3242 are shown in Table 4 and that 
  of NGC\,6369 in Table 5.
     
\begin{table}[h]
\caption[]{Visual spectrum of NGC\,3242.}
\begin{center}
\begin{tabular}{llcccc}
\hline
\hline
\multicolumn{1}{c}{$\lambda$} & Ion & \multicolumn{3}{c}{Intensities$^{\dagger}$}& Average\\ \cline{3-5}
\multicolumn{1}{c}{(\AA)}& & (1) & (2)& (3) & Intens.\\
\hline

3426 & \ion{[Ne}{v]}  &        &  1.73  &       &  1.73 \\
3727$^{\ast}$   & \ion{[O}{ii]} & 14.4 & 6.0  & 11.2  & 12 \\
3869 & \ion{[Ne}{iii]} & 103 & 88.8 & 91  & 93 \\
4101 & \ion{H$\delta$}& 26.6 & 27.0 & 25.8 &     \\
4267 & \ion{C}{ii}    & 0.62   &  0.77 & 0.72  & 0.70 \\
4340 & \ion{H$\gamma$}& 47.4 & 46.8 & 44.6 &     \\
4363 & \ion{[O}{iii]} & 13.5 & 14.0   & 12.0  & 13.5 \\
4471 & \ion{He}{i}    & 4.12 & 3.52 & 3.56 & 3.7 \\
4686 & \ion{He}{ii}  & 25.7 & 40.5 & 23.0 & 28 \\
4711 & \ion{[Ar}{iv]} & 4.89  &  5.95 & 4.0 & 4.9 \\
4724 & \ion{[Ne}{iv]} & 0.0493   &    &     & 0.049\\
4725 & \ion{[Ne}{iv]} & 0.0407  &     &     & 0.041\\
4740 & \ion{[Ar}{iv]} & 4.52    & 5.29 & 3.58 & 4.5 \\
4861 & \ion{H$\beta$} & 100 & 100 & 100 & 100\\
5007 & \ion{[O}{iii]} & 1280 & 1205 & 1260 & 1260\\
5517 & \ion{[Cl}{iii]} & 0.317 & 0.26  &    &  0.30 \\
5538 & \ion{[Cl}{iii]} & 0.277 & 0.25  &      & 0.27 \\
5755 & \ion{[N}{ii]}  & 0.0634  &    & 0.065  & 0.064 \\
5876 & \ion{He}{i}    & 11.6   & 10.3 & 9.2  & 11.0 \\
6312 & \ion{[S}{iii]} & 0.631  & 0.53  & 0.53  & 0.56 \\
6563 & \ion{H$\alpha$}& 285  &       & 282  &      \\
6584 & \ion{[N}{ii]}  & 2.63  & 1.51  & 3.48   & 2.6 \\
6717 & \ion{[S}{ii]}  & 0.278 & 0.24  &      & 0.27 \\
6731 & \ion{[S}{ii]}  & 0.392 & 0.26  &      & 0.39 \\
7005 & \ion{[Ar}{v]}   & 0.0835    &      &  0.063 & 0.080 \\
7135 & \ion{[Ar}{iii]}& 6.99 &     & 6.92  & 6.96 \\
7263 & \ion{[Ar}{iv]}  & 0.124    &   & 0.11  & 0.19: \\
8045 & \ion{[Cl}{iv]}  &      &   & 0.41  & 0.41 \\
C    &                 & 0.10 & 0.10  & 0.14  &  \\
\hline
\end{tabular}
\end{center}

$^{\dagger}$ References; (1) Tsamis et al. (\cite{tsamis}), (2) Krabbe et al.
(\cite{krab}), (3) Aller \& Czyzak (\cite{aller}).\\ 
(:) indicates uncertain values.\\ 
$^{\ast}$ This is a blend of $\lambda$3726 and $\lambda$3729 lines.\\
C is the extinction
\end{table}

\begin{table}[h]
\caption[]{Visual spectrum of NGC\,6369.}
\begin{center}
\begin{tabular}{llcccc}
\hline
\hline
\multicolumn{1}{c}{$\lambda$} & Ion & \multicolumn{3}{c}{Intensities$^{\dagger}$}& Average\\ \cline{3-5}
\multicolumn{1}{c}{(\AA)}& & (1) & (2)& (3) & Intens.\\
\hline

3726 & \ion{[O}{ii]}  &        &    &  63     &  63 \\
3729   & \ion{[O}{ii]} &      &      & 39  & 39 \\
3869 & \ion{[Ne}{iii]} &     & 84.2 & 89.0  & 88 \\
4340 & \ion{H$\gamma$}&     & 46.4 &     &     \\
4363 & \ion{[O}{iii]} &     & 14.2   &  8.3  & 11 \\
4686 & \ion{He}{ii}  &  1.4 & 2.4 & 10.0 & 2.5 \\
4711 & \ion{[Ar}{iv]} &     &      & 1.3   & 1.3 \\
4740 & \ion{[Ar}{iv]} &        &     &  1.5 & 1.5 \\
4861 & \ion{H$\beta$} & 100 & 100 & 100 & 100\\
5007 & \ion{[O}{iii]} & 1241 & 1307 & 1260 & 1260\\
5517 & \ion{[Cl}{iii]} & 1.0 & 0.48  &    &  0.6 \\
5538 & \ion{[Cl}{iii]} & 1.0 & 0.91  &      & 0.96 \\
5755 & \ion{[N}{ii]}  & 2.0  & 1.77   & 1.9  & 1.9 \\
5876 & \ion{He}{i}    & 16   & 15.2 & 15  & 15.5 \\
6312 & \ion{[S}{iii]} & 1.4  & 1.3  & 7.0  & 1.4 \\
6563 & \ion{H$\alpha$}& 287  &  285     &   &      \\
6584 & \ion{[N}{ii]}  & 73  & 62.2  & 87.2   & 73 \\
6717 & \ion{[S}{ii]}  & 4.4 &  2.81    & 4.4  & 4.2 \\
6731 & \ion{[S}{ii]}  & 6.3 &  5.25   &  7.1    & 6.8 \\
7135 & \ion{[Ar}{iii]}&    &  16.6   &     & 16.6 \\

\hline
\end{tabular}
\end{center}

$^{\dagger}$ References; (1) Monteiro et al. (\cite{mont}), (2) Aller \& Keyes
(\cite{allerkey}), (3) Pena et al. (\cite{pena}).\\ 
(:) indicates uncertain values.\\ 
$^{\ast}$ This is a blend of $\lambda$3726 and $\lambda$3729 lines.
\end{table}

  \subsection{The IUE ultraviolet spectrum }

  Only a few IUE observations of NGC\,6369 were made. All of them were
  underexposed in spite of the fact that the nebula is so bright. The
  very large extinction reduces the ultraviolet emission in this case
  to such an extent that the measurements are not useful. By contrast
  there are 91 low resolution {\em IUE} observations of NGC\,3242 as
  well as six high resolution observations of this nebula. Seventy of
  the low resolution measurements were taken with the large aperture
  (10\arcsec x 23\arcsec) with varying exposure times, and 12 were
  taken with a small aperture (3\arcsec~diameter).  We find that the
  small aperture measurements are too noisy and we do not use them.
  The large aperture measurements do not cover the entire nebula.  We
  have used two of the highest S/N observations made with long
  exposure times both in the short and the long wavelength regions.
  These are SWP15495 and SWP16418 for the short wavelength region and
  LWR11973 and LWR 12678 for the long wavelength region. We have
  checked that the strongest lines are not saturated by measuring
  these same lines on spectra with shorter exposure times: SWP15495,
  SWP 16419 and LWR12679. In addition two high resolution large
  aperture measurements are used: SWP03643 and LWR03206. All of these
  measurements are centered within 6\arcsec of the center of the
  nebula and thus are representative of the nebula as a whole.  The
  results are shown in Table 6. The uncertainties are about the same
  as given by Henry et al.  (\cite{henry}): about 10\% for the
  strongest lines and about 25\% for the weaker lines. From the high
  resolution spectra we find the \ion{C}{iii} ratio
  1906/1909=1.4$\pm$0.1 and the \ion{Ne}{iv} ratio
  2422/2425=0.92$\pm$0.5.

  The extinction correction was made by assuming a theoretical ratio
  for the \ion{He}{ii} line ratio $\lambda$1640/$\lambda$4686\,\AA~ at
  T=11\,500 K and an N$_e$ of 10$^4$ cm$^{-3}$. The ratio of
  $\lambda$1640 to H$\beta$ can then be found using the
  $\lambda$4686/H$\beta$ ratio in Table 4.  A further correction for
  extinction relative to $\lambda$1640\AA~is then made using the
  reddening curve of Fluks et al.(\cite{fluks}) but because of the
  small extinction this is never more than 20\%, which indicates the
  uncertainties of the UV intensities above the errors of measurement
  given above. The results are shown in the last two columns of Table
  6.

\begin{table}[htbp]
\caption[]{IUE Spectrum of NGC\,3242.}  
\begin{center}
\begin{tabular}{llccc}
\hline
\hline
\multicolumn{1}{c}{$\lambda$} & Ion &\multicolumn{3}{c}{Intensities}\\
\cline{3-5}
\multicolumn{1}{c}{(\AA)}& & (1) & (2) &  (I/H$\beta$)   \\
\hline

1239 & \ion{N}{v}    &  19   & 1.32  & 7.4 \\
1335 & \ion{C}{ii}   & 16  & 1.0  &  4.76 \\
1400 & \ion{O}{iv}  &  7.8  & 4.68 & 2.19 \\
1485 & \ion{N}{iv]} &  26  & 15.6  & 6.44 \\
1548 & \ion{C}{iv}  &  110  & 66.0  & 30.8  \\
1575 & \ion{[Ne}{v]} & 12   &  7.2  &  3.36 \\
1640 & \ion{He}{ii}  & 622   & 373  & 174  \\
1663 & \ion{O}{iii]} & 45   & 27 & 12.6 \\
1750 & \ion{N}{iii]} & 15.8  &  9.48 & 4.43 \\
1909 & \ion{C}{iii]} & 675    & 405  & 189  \\
2298 &\ion{C}{iii]} & 40    &  27   & 12.6  \\
2325 & \ion{C}{ii]}  & 26.8   & 17.0 & 7.95 \\
2423 & \ion{[Ne}{iv]} & 44   & 30.1  & 14.1  \\
2512 & \ion{He}{ii}  &  17  & 8.67 & 4.05 \\
2733 & \ion{He}{ii}  & 27  &  11.4 & 5.32  \\
2837 & \ion{[Fe}{iv]}  & 45 & 22.2 & 10.4  \\
3024 & \ion{O}{iii}  &  20  & 7.8  & 3.64  \\
3048 & \ion{O}{iii}  &  50  &  19.8 & 9.24  \\
3133 & \ion{O}{iii}  & 100  &  39.0 & 18.2  \\
3103  & \ion{He}{ii}  & 38  &  13.8 &  6.44 \\

\hline 
\end{tabular}
\end{center}
(1)Measured intensity from low resolution spectra in 
units of 10$^{-13}$ erg cm$^{-2}$ s$^{-1}$. \\
(2)Intensity corrected for diaphragm size and extinction in units of
10$^{-12}$ erg cm$^{-2}$ s$^{-1}$. 
I/H$\beta$ is normalized to H$\beta$=100.\\ 
\end{table}

\section{Chemical composition of the nebulae}

The method of analysis is as follows.. First the electron density and 
temperature as a function
of the ionization potential are determined. The ionic abundances
are then determined, using density and temperature appropriate for the ion
under consideration. Then the element abundances are found for those
elements in which a sufficient number of ionic abundances have been
derived.

\subsection{Electron density}  

The ions used to determine $N_{\mathrm{e}}$ for NGC\,3242 are listed
in the first column of Table\,7; those for NGC\,6369 are listed in
Table 8. The ionization potential required to reach this stage of
ionization, and the wavelengths of the lines used, are given in
Cols.\,2 and 3 of the tables. Note that the wavelength units are
\AA~when 4 figures are given and microns when 3 are shown. The
observed ratio of the lines is given in the fourth column; the
corresponding $N_{\mathrm{e}}$ is given in the fifth column. The
temperature used is discussed in the following section, but is
unimportant since these line ratios are essentially determined by the
density. The density from the \ion{C}{iii]} lines in NGC\,3242 is
uncertain because the two measurements used differed substantially..

There is no indication that the electron density varies with
ionization potential in a systematic way in either of the nebulae.
The electron density appears to be about 2500 cm$^{-3}$ in NGC\,3242
and slightly higher (about 2800 cm$^{-3}$) in NGC\,6369. The error is
about 20\% in NGC\,3242 and slightly higher in NGC\,6369. We will use
these densities in further discussion of the abundances, but any value
between 1500 cm$^{-3}$ and 3000 cm$^{-3}$ will give the same values of
abundance.

\begin{table}[t]
\caption[]{ Electron density indicators in NGC\,3242.}
\begin{center}
\begin{tabular}{lcccc}
\hline
\hline
Ion &Ioniz. & Lines& Observed &N$_{\mathrm{e}}$ \\
&Pot.(eV) & Used  & Ratio & (cm$^{-3}$)\\
\hline
\ion{[S}{ii]} & 10.4 & 6731/6716 & 1.38  & 1900\\
\ion{[O}{ii]} & 13.6 & 3626/3729 & 1.61 & 2700 \\
\ion{[S}{iii]} & 23.3 & 33.5/18.7 & 0.38: & 2800 \\
\ion{[Cl}{iii]} & 23.8 & 5538/5518 & 1.08 & 3000\\
\ion{C}{iii]} & 24.4 & 1906/1909 & 1.29 & 3300:\\
\ion{[O}{iii]} & 35.1 & 51.8/88.4 & 2.24 & 900:\\
\ion{[Ar}{iv]} & 40.7 & 4711/4740 & 1.08 & 2100:\\ 
\ion{[Ne}{iv]} & 63.5 & 2425/2422 & 0.92 & 5000: \\

\hline
\end{tabular}
\end{center}
: Indicates uncertain values.\\ Wavelengths of the far infrared lines are in
microns, the rest are in Angstroms.

\end{table}

\begin{table}[t]
\caption[]{ Electron density indicators in NGC\,6369.}
\begin{center}
\begin{tabular}{lcccc}
\hline
\hline
Ion &Ioniz. & Lines& Observed &N$_{\mathrm{e}}$ \\
&Pot.(eV) & Used  & Ratio & (cm$^{-3}$)\\
\hline
\ion{[S}{ii]} & 10.4 & 6731/6716 & 1.51  & 2300\\
\ion{[O}{ii]} & 13.6 & 3626/3729 & 1.62 & 2800 \\
\ion{[S}{iii]} & 23.3 & 33.5/18.7 & 0.409 & 2900 \\
\ion{[Cl}{iii]} & 23.8 & 5538/5518 & 1.3: & 3000:\\
\ion{[O}{iii]} & 35.1 & 51.8/88.4 & 2.1 & 800:\\
\ion{[Ar}{iv]} & 40.7 & 4711/4740 & 0.87 & 4000:\\

\hline
\end{tabular}
\end{center}
: Indicates uncertain values. Wavelength units as in Table 7.

\end{table}

  \subsection{Electron temperature}

  A number of ions have lines originating from energy levels far
  enough apart that their ratio is sensitive to the electron
  temperature. These are listed in Tables 9 and 10 for each of the
  nebulae. These tables are arranged similarly to the previous tables.
  No temperature gradient is seen in either of the nebulae. Both
  nebulae have very similar electron temperatures: the temperature of
  NGC\,3242 is about 11,500 K while that of NGC\,6369 seems to be
  slightly lower: T=11,000 K. The temperature for NGC\,3242 is better
  determined because ultraviolet measurements are available. Although
  a weak line is seen at 14.3$\mu$m, no temperature can be obtained
  from \ion{Ne}{v} because no good measurement of the ultraviolet line
  at $\lambda$3425 has been made.


\begin{table}[t]
\caption[]{Electron temperature indicators in NGC\,3242.}
\begin{center}
\begin{tabular}{lcccc}
\hline
\hline
 Ion & Ioniz. & Lines& Observed & $T_{\mathrm{e}}$\\
 & Pot.(eV)& Used$\dagger$ &Ratio  & (K) \\
\hline

\ion{[N}{ii]}  & 14.5 & 5755/6584 & 0.0211 & 11\,000 \\
\ion{[S}{iii]} & 23.3 & 6312/18.7 & 0.0.075 & 11\,800 \\
\ion{[Ar}{iii]} & 27.6 & 7136/21.8 & 20.3 & 10\,600\\
\ion{[O}{iii]} & 35.1 & 4363/5007 & 0.0106 & 11\,780\\
\ion{[O}{iii]} & 35.1 & 1663/5007  & 0.0099 & 10\,800\\
\ion{[O}{iii]} & 35.1 & 5007/51.8  & 7.90 & 11\,000\\
\ion{[Ne}{iii]} & 41.0 & 3869/15.5 & 0.89 & 11\,000\\
\ion{[O}{iv]}   & 54.9 & 1400/25.9 & 0.016 & 12\,000:\\

\hline
\end{tabular}
\end{center}
: Indicates uncertain value.\\ 
$\dagger$When the wavelength has 4 figures it has the 
units of Angstrom and 3 figures is micron.
\end{table}


\begin{table}[t]
\caption[]{Electron temperature indicators in NGC\,6369.}
\begin{center}
\begin{tabular}{lcccc}
\hline
\hline
 Ion & Ioniz. & Lines& Observed & $T_{\mathrm{e}}$\\
 & Pot.(eV)& Used$\dagger$ &Ratio  & (K) \\
\hline

\ion{[N}{ii]}  & 14.5 & 5755/6584 & 0.022 & 11\,000 \\
\ion{[S}{iii]} & 23.3 & 6312/18.7 & 0.0.049 & 10\,000 \\
\ion{[Ar}{iii]} & 27.6 & 7136/8.99 & 1.07 & 10\,000\\
\ion{[O}{iii]} & 35.1 & 4363/5007 & 0.0096 & 11\,300\\
\ion{[O}{iii]} & 35.1 & 5007/51.8  & 11.3 & 11\,500\\
\ion{[Ne}{iii]} & 41.0 & 3869/15.5 & 0.604 & 10\,500\\

\hline
\end{tabular}
\end{center}
: Indicates uncertain value.\\ 
$\dagger$When the wavelength has 4 figures it has the 
units of Angstrom and 3 figures is micron.
\end{table}

  \subsection{Ion and element abundances}

  The ion abundances have been determined using the following
  equation:

  \begin{equation}
  \frac{N_{\mathrm{ion}}}{N_{\mathrm{p}}}= \frac{I_{\mathrm{ion}}}{I_{\mathrm{H_{\beta}}
  }} N_{\mathrm{e}}
  \frac{\lambda_{\mathrm{ul}}}{\lambda_{\mathrm{H_{\beta}}}} \frac{\alpha_{\mathrm{H_{\beta}}}}{A_{\mathrm{ul}}}
  \left( \frac{N_{\mathrm{u}}}{N_{\mathrm{ion}}} \right)^{-1} 
  \label{eq_abun}
  \end{equation}

  where $I_{\mathrm{ion}}$/$I_{\mathrm{H_{\beta}}}$ is the measured
  intensity of the ionic line compared to H$\beta$, $N_{\mathrm{p}}$
  is the density of ionized hydrogen, $\lambda_{\mathrm{ul}}$ is the
  wavelength of the line, $\lambda_{\mathrm{H_\beta}}$ is the
  wavelength of H$\beta$, ${\alpha_{\mathrm{H_\beta}}}$ is the
  effective recombination coefficient for H$\beta$, $A_{\mathrm{ul}}$
  is the Einstein spontaneous transition rate for the line, and
  $N_{\mathrm{u}}$/$N_{\mathrm{ion}}$ is the ratio of the population
  of the level from which the line originates to the total population
  of the ion. This ratio has been determined using a five level atom.
  The atomic data used is given in the paper of Pottasch \& Beintema
  (\cite{pott1}).  A 5 level atom is used in all cases. For iron this is
  justified because collisional rates to higher levels followed by
  cascade through the lower levels are quite small in comparison to
  direct collisions to the lower levels.

  The results are given in Tables 11 and 12, where the first column
  lists the ion concerned, the second column the line used for the
  abundance determination, and the third column gives the intensity of
  the line used relative to H$\beta$=100. The fourth column gives the
  value of the ionic abundance assuming the ion is formed at T=11\,500
  K for NGC\,3242 and T=11\,000 K for NGC\,6396, while the fifth
  column gives the ionization correction factor (ICF), which has been
  determined empirically. Notice that the ICF is close to unity for
  all elements listed in the table except for Fe and P.

  The error of measurement of the IRS intensities as can be seen in
  Table 1 is usually small, often not more than 5-7\%. In the few
  cases when the error is large this has either been indicated with a
  ':' or by not using the line.  The correction for adjusting the SH
  to LH intensity scales and the diaphragm size is also small, about
  10\%. This includes the assumption that the unmeasured parts of the
  nebula have the same composition as the measured parts. This has
  been checked in the optical region by Barker (\cite{barker}) and can
  be seen to be a reasonable approximation in the ultraviolet region
  by comparing the {\em IUE} spectra obtained in different parts of
  the nebula. The uncertainty of the collisional strengths introduces
  an error of 10-15\% so that the total error for the ions of neon,
  sulfur and argon determined with the IRS measurements is less than
  20\%. This will also be true of the abundances of these elements
  because the ICF for these elements is close to unity. The error for
  the nitrogen and oxygen abundances is somewhat higher because the
  visual and ultraviolet measurements are less certain. In addition
  the temperature is more important for these ions and the total
  errors may be twice as large. The element abundances are given in
  the last column. The carbon recombination line abundance for
  NGC\,3242 is given at the end of Table 11. It is somewhat more than
  a factor of 3 higher than the value obtained from the collisional
  transition. This difference is found in other PNe as well but is not
  yet understood. No recombination line has been measured in
  NGC\,6369.

  The helium abundance was derived using the theoretical work of
  Benjamin et al. (\cite{benjamin}) and Porter et
  al. (\cite{porter}) For recombination of singly ionized helium,
  most weight is given to the $\lambda$ 5875\AA~line, because the
  theoretical determination of this line is the most reliable.

\begin{table}[h]
  \caption[]{Ionic concentrations and chemical abundances in NGC\,3242.
    Wavelengths in Angstrom for all values of $\lambda$ above 1000, otherwise
    in $\mu$m.}


\begin{tabular}{lccccc}
\hline
\hline
Ion & $\lambda$ &  (I/H$\beta$)$\dagger$  & $N_{\mathrm{ion}}$/$N_{\mathrm{p}}$ 
& ICF & $N_{\mathrm{el.}}$/$N_{\mathrm{p}}$\\
\hline

He$^{+}$  & 5875 &  11.1   & 0.0707    &     &  \\
He$^{++}$ & 4686 &  25     & 0.0212    & 1.0 & 0.092 \\
C$^{+}$   & 2324 & 7.65    & 6.8(-6)  &     &    \\
C$^{++}$  & 1909 & 181     & 1.61(-4) &     &   \\
C$^{+3}$  & 1548 & 29.4    & 2.6(-5)  & 1.0 & 1.95(-4) \\
N$^{+}$   & 6584 & 2.7     & 4.1(-7) &     &   \\
N$^{++}$  & 1750 & 4.23    &   (-5) &     &    \\
N$^{++}$  & 57.3 & 14.5    & 5.7(-5) &     &    \\
N$^{+3}$  & 1485 & 7.0     & 4.39(-5) &     &     \\
N$^{+4}$  & 1239 & 5.1:    & 3.6(-5): & 1.0  & 1.35(-4)   \\
O$^{+}$   & 3729 & 5.36    & 5.1(-6) &     &     \\
O$^{++}$  & 5007 & 1280    & 3.18(-4) &     &    \\
O$^{++}$  & 51.7 & 162     & 4.38(-4)  &     &     \\
O$^{+3}$  & 25.8 & 137     & 4.3(-5)  & 1.0 & 3.8(-4) \\
Ne$^{+}$  & 12.8 & 1.31    & 1.8(-6) &     &       \\
Ne$^{++}$ & 15.5 & 116     & 7.53(-5) &     &       \\
Ne$^{++}$ & 3869 & 103     & 6.5(-5) &     &    \\
Ne$^{+3}$ & 2423 & 12.6    & 1.25(-5) &     &    \\
Ne$^{+4}$ & 14.3 & 0.31    & 3.1(-8) & 1.0 & 9.0(-5)  \\  
S$^{+}$   & 6731 & 0.392    & 1.8(-8)  &     &      \\
S$^{++}$  & 18.7 & 8.33     & 8.38(-7) &     & \\
S$^{++}$  & 6312 & 0.61    & 8.3(-7)  &     &     \\
S$^{+3}$  & 10.5 & 70.3    & 1.94(-6) & 1.0 & 2.8(-6) \\
Ar$^{++}$ & 8.99 & 8.48    & 8.32(-7) &     &       \\
Ar$^{++}$ & 21.8 & 0.345   & 5.2(-7) &     &       \\
Ar$^{++}$ & 7135 & 7.0    & 5.0(-6) &     &         \\
Ar$^{+3}$ & 4740 & 4.3    & 8.49(-7) &     &       \\
Ar$^{+4}$ & 7005 & 0.083  & 1.4(-8)  & 1.0 & 1.7(-6) \\
Cl$^{++}$ & 5538 & 0.28    & 2.7(-8)  &     &       \\ 
Cl$^{+3}$ & 11.8 & 0.616   & 3.51(-8) &     &      \\
Cl$^{+3}$ & 8045 & 0.5:    & 4.0(-8): & 1.1 & 7.5(-8) \\
Fe$^{++}$ & 22.9 & 1.55   & 4.7(-7) &     &      : \\ 
Fe$^{++}$ & 33.0 & 0.51   & 5.8 (-7) & 1.6 & 8.0(-7) \\
P$^{++}$  & 17.9 & 0.322   & 1.33(-8): & 3.5:& 4.5(-8):  \\
C$^{++}$  & 4267 & 0.62    & 6.0(-4):  &     &  \\
 
\hline
\end{tabular}

$\dagger$Intensities given with respect to H$\beta$=100.\\
: Indicates uncertain value.

\end{table}

\begin{table}[h]
  \caption[]{Ionic concentrations and chemical abundances in NGC\,6369.
    Wavelengths in Angstrom for all values of $\lambda$ above 1000, otherwise
    in $\mu$m.}


\begin{tabular}{lccccc}
\hline
\hline
Ion & $\lambda$ &  (I/H$\beta$)$\dagger$  & $N_{\mathrm{ion}}$/$N_{\mathrm{p}}$ 
& ICF & $N_{\mathrm{el.}}$/$N_{\mathrm{p}}$\\
\hline

He$^{+}$  & 5875 &  15.5   & 0.100    &     &  \\
He$^{++}$ & 4686 &  2.2     & 0.0018    & 1.0 & 0.102 \\
N$^{+}$   & 6584 & 73     & 1.17(-5) &     &   \\
N$^{+}$  & 122 & 0.33    &  1.85 (-5) &     &    \\
N$^{++}$  & 57.3 & 16.9    & 6.57(-5) &  1.2   & 9.4 \\
O$^{+}$   & 3729 & 39    & 4.2(-5) &     &     \\
O$^{++}$  & 5007 & 1260    & 3.42(-4) &     &    \\
O$^{++}$  & 51.8 & 111     & 3.0(-4)  &     &     \\
O$^{+3}$  & 25.8 & 10.6     & 3.2(-6)  & 1.0 & 3.9(-4) \\
Ne$^{+}$  & 12.8 & 4.1    & 6.25(-6) &     &       \\
Ne$^{++}$ & 15.5 & 142     & 8.97(-5) &     &       \\
Ne$^{++}$ & 3869 & 88     & 5.29(-5) & 1.0    & 9.7(=5) \\  
S$^{+}$   & 6731 & 6.8   & 3.22(-7)  &     &      \\
S$^{++}$  & 18.7 & 28.6     & 2.83(-6) &     & \\
S$^{++}$  & 6312 & 1.4   & 2.13(-6)  &     &     \\
S$^{+3}$  & 10.5 & 90    & 2.40(-6) & 1.0 & 6.0(-6) \\
Ar$^{++}$ & 8.99 & 15.6    & 1.48(-6) &     &       \\
Ar$^{++}$ & 7135 & 16.6    & 1.25(-6) &     &         \\
Ar$^{+3}$ & 4740 & 1.5   & 3.22(-7) & 1.0    & 1.6(-6) \\
Cl$^{+}$ &  14.3 & 0.060 & 1.1(-8)  &       &        \\
Cl$^{++}$ & 5538 & 0.95    & 1.01(-7)  &     &       \\ 
Cl$^{+3}$ & 11.8 & 0.308   & 1.77(-8) & 1.0    & 1.2(-7) \\
P$^{++}$  & 17.9 & 0.814   & 3.42(-8): & 2.0:& 7.0(-8):  \\

\hline
\end{tabular}

$\dagger$Intensities given with respect to H$\beta$=100.\\
: Indicates uncertain value.

\end{table}

\section{Comparison with other abundance determinations} 
 
Tables 13 and 14 show a comparison of our abundances with three of the
most important determinations in the past 20 years for each nebulae.
For NGC\,3242 reasonable agreement is found. Good agreement is found
for oxygen; this is because the same electron temperature is used for
the most important oxygen ion. For the other elements the agreement is
less good. For nitrogen a somewhat higher abundance is found which is
probably due to the fact that the most important ionization stages
have not been measured earlier. Nitrogen is somewhat higher than in
the Sun. A C/O ratio substantially lower than unity and close to the
solar value is found but the error is about 30\%. Sulfur is in good
agreement with earlier determinations but is a factor of 4 lower than
the solar value. This is a general phenomenon in PNe and has been
discussed earlier (e.g. Pottasch \& Bernard-Salas \cite{pbs},
Bernard-Salas et al. \cite{bsp}).  Phosphorus, which has never been
measured before in this nebula and is slightly uncertain because only
a single ionization stage has been measured, is considerably lower
than solar. Chlorine also seems rather strongly depleted compared to
the solar abundance. The solar abundance listed in the table is taken
from Asplund et al. (\cite{asplund}).  Note that for solar sulfur and
chlorine more weight has been given to the abundance determination in
meteorites since this determination is more accurate than for the Sun
itself.  Neon and argon abundances are taken from the references given
in Pottasch and Bernard-Salas (\cite{pbs}) and differ substantially
with those given by Asplund et al. (\cite{asplund}). We point out that the
solar abundances have changed greatly in the past 10 years and it is uncertain
that there will be no further changes in the near future (see discussion by
Bernard-Salas et al.\cite{bsp}).


\begin{table}[htbp]
\caption[]{Comparison of abundances in \object{NGC\,3242}.}
\begin{tabular}{lrrrrr}
\hline
\hline
Elem.  & Present & T$^{\dagger}$ & H$^{\dagger}$ & B$^{\dagger}$ & Solar$^{\dagger}$  \\ 
\hline  

He & 0.092 & 0.10  & 0.081 & 0.091 &   0.098   \\
C(-4)  & 1.95 &    & 4.12 & 2.6  &   2.5  \\
N(-4)  & 1.35 & 0.34  & 0.61  & 0.89 &  0.84 \\
O(-4)  & 3.8 & 3.31 & 3.38  &  4.35    &  4.6  \\
Ne(-5) & 9.0  & 7.8 & 5.1 & 11.0  & 12   \\
S(-6)  & 2.8  & 2.38 &     & 3.2  &  14   \\   
Ar(-6) & 1.7 & 0.98 &      & 1.4 &   4.2 \\
Cl(-7) & 0.70 & 0.88 &     &     & 3.5   \\
P(-8)  & 4.5 &    &      &     &  23   \\

\hline  

\end{tabular}

$^{\dagger}$References: T: Tsamis et al. (\cite{tsamis}), H: Henry et al. 
(\cite{henry}), B: 
Barker (\cite{barker}),  Solar: Asplund et al. (\cite{asplund}), except Ne and 
Ar (see Pottasch \& Bernard-Salas ,\cite{pbs}).

\end{table}

For NGC\,6369 the agreement is somewhat less good. The reason for this is that
the comparison determinations use only the visible spectra since no ultraviolet
measurements have been made. For nitrogen only the N$^{+}$ abundance can be 
measured and a large correction for higher ionization stages must be made. 
Besides this the very high extinction correction which must be made in this
nebula leads to larger uncertainties in line intensities. The advantage in
using the infrared measurements is quite clear.

\begin{table}[htbp]
\caption[]{Comparison of abundances in \object{NGC\,6369}.}
\begin{tabular}{lrrrrr}
\hline
\hline
Elem.  & Present & M$^{\dagger}$ & P$^{\dagger}$ & AK$^{\dagger}$ & Solar$^{\dagger}$  \\ 
\hline  

He(-1) & 1.02 & 1.14  & 1.14 & 1.3 &   0.98   \\
N(-5)  & 7.9 & 10  & 2.2  & 12 &  8.4 \\
O(-4)  & 4.0 & 6.1 & 4.7  &  2.9    &  4.6  \\
Ne(-5) & 9.7  & 6.5 & 9.4 & 4.0  & 12   \\
S(-6)  & 6.0  & 7.5 &     & 6.0  &  14   \\   
Ar(-6) & 1.6 &      &      & 4.2 &   4.2 \\
Cl(-7) & 1.2 &      &     & 2.6   & 3.5   \\
P(-8)  & 7.0 &      &      &      &  23   \\

\hline  

\end{tabular}

$^{\dagger}$References: M:Monteiro et al. (\cite{mont}), P:Pena et al. 
(\cite{pena}), AK:Aller \& Keyes (\cite{allerkey}),  Solar: Asplund et al. 
(\cite{asplund}), except Ne and Ar (see Pottasch \& Bernard-Salas, \cite{pbs}).

\end{table}

\section{Discussion of stellar evolution}

For both nebulae there is a general agreement with the (more uncertain) 
earlier values. Helium, oxygen and nitrogen abundances are essentially
solar. This is probably true of neon as well. On the other hand,
sulfur, argon, chlorine and phosphorus are all significantly lower
than solar by a factor of 3 to 4, both in NGC\,6369 and NGC\,3242.
This is also remarkably similar to most elliptical PNe which have been
studied in the infrared and which are at approximately the same
galactocentric distance. The abundances of these nebulae are given in
Table 15.  These nebulae not only have similar morphology, but are
also all at rather high galactic latitude. The last five PNe listed
all have galactic latitudes between 10\degr and 20\degr, which is much
higher than the average value. Part of this difference occurs because
these are all relatively nearby nebulae. But the main reason for the
high galactic latitude is probably that these nebulae are formed from
stars which initially had a relatively low mass.

\begin{table}[htbp]
\caption[]{Comparison of abundances in several elliptical PNe.}
\begin{tabular}{lrrrrrrr}
  \hline
\hline
Elem.  & 3242$^{\dagger}$ & 6369$^{\dagger}$ & 2392$^{\dagger}$ & 6826$^{\dagger}$ & 7662$^{\dagger}$ & 418$^{\dagger}$ & 2022$^{\dagger}$  \\ 
\hline  

He(-1)  & 0.92 & 1.02  & 0.80 & 0.95 & 0.88 &      & 1.06  \\
C(-4)  & 1.95  &        & 3.3   & 5.3   &  3.6  & 6.2  &  3.7   \\
N(-4)  & 1.35  & 0.79   & 1.85  & 0.50 &  0.67  & 0.95 &  0.99  \\
O(-4)  & 3.8   & 4.0    & 2.9   &  4.0  &  4.2  & 3.5  &  4.7  \\
Ne(-5) & 9.0   & 9.7    & 8.5   & 15    & 6.4   & 8.8  &  13.4   \\
S(-6)  & 2.8   & 6.0    & 5.0   & 2.8   & 6.6   & 4.4  &  6.3   \\   
Ar(-6) & 1.7   & 1.6    & 2.2   & 1.4   & 2.1   & 1.8  &  2.7   \\
Cl(-7) & 0.70 & 1.2     & 1.3   & 0.9   &       & 1.2  &  1.3   \\
P(-8)  & 4.5   & 7.0    & 6.5   & 17    &       &      &      \\
Fe(-7) &       &        & 8.0   & 6.1   &       &      &       \\

\hline  

\end{tabular}

$^{\dagger}$References: NGC\,3242 and 6369: this paper, NGC\,2392: Pottasch et 
al. (\cite{pott8}), NGC\,6826: Surendiranath et al. (\cite{nath}), NGC\,7662:
Pottasch et al. (\cite{pott4}), IC\,418: Pottasch et al. (\cite{pott6}), 
NGC\,2022: Pottasch et al. (\cite{pott7}).

\end{table}

In considering all the nebulae in Table 15 the following is noted.
The low helium abundance indicates that no helium has been produced
which implies that the second dredge-up and hot-bottom burning have
not taken place. For some PNe the carbon abundance is somewhat higher
than oxygen suggesting that the third dredge-up has taken place in these 
nebulae. Following the models of Karakas (\cite{karakas}) for the case
Z=0.008 (which is close to the solar abundance) the increase in the
carbon abundance will occur at a stellar mass of about 1.7M\smallsun.
The lower mass models for which the carbon has not yet increased show
an increased nitrogen abundance, sometimes by a factor of two or three.
When the increased carbon is found the nitrogen is predicted to have
its original (solar) value. This seems to be exactly what is observed.
Of the nebulae listed in Table 15 we therefore predict that NGC\,3242,
NGC\,7662 and NGC\,2022 have descended from stars with a mass of
between 0.8M\smallsun~and 1.5M\smallsun~and NGC\,6826, NGC\,2392 and
IC\,418 are descendents of stars from between 1.6M\smallsun and
1.9M\smallsun. Above this mass Karakas predicts an increased neon abundance
by more than a factor of two, which is not seen. NGC\,6369 is harder to 
classify since no carbon has yet been measured. From the low nitrogen abundance
we tentatively conclude that this nebula is probably in the second (high mass) 
group.

Notice that for all the nebulae listed in Table 15 the elements S, Ar,
Cl, P and Fe are underabundant compared to the Sun. There are several
possible reasons for this. First the solar abundances may be wrong.
Secondly our determinations of the abundances may be wrong. Thirdly
the elements may be tied up in the dust grains in the nebula. We do
not believe that our determination of the abundances can be so badly
wrong. The worst case is P because only a single ionization stage is
observed and a reasonably large correction for the missing stages of
ionization has been made. It seems unlikely that this has been so
poorly done. For S, Ar and Cl the solar abundance is somewhat
uncertain and could be wrong. This is certainly the case for Ar for
which a coronal value must be used. Furthermore the difference between
the nebular value and the solar value is only a factor of two to three
for these 3 elements. For S and Cl it is not possible
to say whether the difference is due to wrong solar values or that the
elements are in taken up in the dust. Ar is a noble gas and it is unlikely that
it can be present in dust. For P and Fe the difference with
the solar value is so large that these elements must be in the form of
dust. Since iron has been known to be depleted in all PNe in which it has
been observed, there has been some speculation in the literature as to 
whether iron molecules have been observed in the infrared continuum. Hony et
al.\,(\cite{hony}) claim to have detected FeS in two PNe. These authors
also note that iron sulfides are abundantly found in meteoritic material.
Iron could also be in the form of FeO. But iron molecules are easily
destroyed. Iron is highly refractory and should easily stick to carbonaceous 
dust grains which are present in PNe and are responsible for their IR
continuum. Measuring the iron content of the dust grains is very difficult
and has not yet been done. But it is significant that about 1\% of the nebular 
mass must be contained in dust containing this iron molecule assuming that
the original composition was solar. Notice that the spectrum of dust does not
show any hydrocarbon (PAH) emission bands in the oxygen-rich nebulae listed 
in Table 15. Very weak PAH features are seen in the carbon-rich nebula
IC\,418 and in NGC\,6369.

\section{The central star}

  \subsection{Stellar temperature}

  As discussed in the introduction, the spectrum of the central star
  of NGC\,3242 has been studied by several authors. Pauldrach et al.
  (\cite{paul}) have fitted model atmospheres to the measured
  ultraviolet stellar spectrum and Kudritzki et al. (\cite{kud}) have
  studied the optical spectrum.  Pauldrach et al.(\cite{paul})
  determined the effective stellar temperature from the FeIV/FeV
  ionization balance to be 75\,000 K while Kudritzki et al.
  (\cite{kud}) using the ionization equilibrium of HeI and HeII also
  obtained a value of 75\,000 K. Neither of these authors has
  studied the spectrum of the central star of NGC\,6369. This star is
  of the Wolf-Rayet type and was originally classified as WC4 (see
  Acker et al. \cite{acker}).  More recently it is classified as WO3
  (Acker \& Neiner \cite{ackern}). The stellar temperature
  associated with this spectral type is usually about 85,000 K to
  90,000 K, which is often determined by the Zanstra temperature\,
  (Acker \& Neiner \cite{ackern}). This at first sight it would
  appear that the central star of NGC\,6369 has the higher temperature
  of the two. The Zanstra temperatures give a slightly different
  picture. The hydrogen Zanstra temperature
  $T_{\mathrm{z}}$(H)$=$69,000 K for NGC\,6369 and 57,000 K for
  NGC\,3242 while the ionized helium Zanstra temperature is 71,000 K
  for NGC\,6369 and 90,000 K for NGC\,3242. Since the ionized helium
  Zanstra temperature is often taken as representative of the actual
  temperature this would imply that NGC\,3242 is the hotter star. This
  is a direct result of the much stronger \ion{He}{ii} line in the
  latter nebula. This is also reflected in the ionic distribution of
  the other elements. For example, 10\% of the oxygen is in the form
  of O$^{+3}$ in NGC\,3242 while only 1\% of the oxygen is in this
  ionization stage in NGC\,6369.
  
  The stellar temperature can also be estimated from the energy
  balance temperature of the central star. This method makes use of the fact
  that the average excess energy per ionizing photon (hence the temperature)
  can be found from the ratio of collisionally excited lines to H$\beta$. 
  For NGC\,3242 this can be found from the sum of the values given in Tables 
  1, 4, and 6. Unobserved lines must be  estimated usually from predicted 
  ratios of observed lines. For NGC\,6369 it is the sum of values given in 
  Tables 3 and 5 again corrected for unobserved lines.. For this latter nebula 
  no ultraviolet measurements are available. For NGC\,3242 the
  ultraviolet measurements are about 20\% of the total and because the
  spectra of the two nebulae are so similar we will assume this to be
  the same in NGC\,6369.  This value of the ratio of the summation
  I/H$\beta$ is about 25 or 26 in both cases.  This could be slightly
  higher if there are important unmeasured lines.  To convert this
  value to a stellar temperature it is not possible to use a simple equation
  because the result, at least for higher temperatures, depends on the amount
  of helium present in the nebula, its ionization state, the opacity in
  the nebula and ultraviolet spectral distribution of the exciting star. To
  obtain a temperature the formulation of Preite-Martinez
  \& Pottasch (\cite{pmp}) is used, assuming blackbody radiation from
  the central star. This leads to a value for the energy balance
  temperature ($T_{\mathrm{z}}$(EB)=70,000 K in both cases. If a model
  atmosphere had been used instead of a blackbody, the energy balance
  temperature could be lower.  When the three values of temperature
  are compared for NGC\,6369 they are consistent and we adopt a
  temperature T=70,000 K for the central star of NGC\,6369. Because of
  the higher ionization state in NGC\,3242, the temperature of the
  central star of this nebula must be somewhat higher since it cannot
  be explained by a lower density. A temperature T=80,000 K is used because
  of the higher ionized helium Zanstra temperature..

  If the distance to these nebulae is known the radius and luminosity
  of the central star may be found. For NGC\,3242 an expansion
  distance has been measured by Hajian et al. (\cite{hajian}) and
  corrected by Mellema (\cite{mell}) to a somewhat uncertain value d=550 pc. 
  The distance to NGC\,6369 has only been determined stastisically and thus is
  quite uncertain; we shall use a value of d=1
  kpc which is an average value of the statistical distances listed in
  Acker et al. (\cite{acker}). The nebula could be somewhat closer
  because it is the third brightest PNe after correction for
  extinction.  The visual magnitudes for the central stars are
  m$_v$=12.43 and 15.91 for NGC\,3242 and NGC\,6369 respectively. When
  corrected for the extinction found in Sec.\,3 these values are
  m$_v$=11.86 and 11.42, which leads to radii of 0.141R\smallsun~and
  0.336R\smallsun~respectively. Combined with the above values of
  central star temperature, the luminosities of these stars are
  730L\smallsun~and 2430L\smallsun~for NGC\,3242 and NGC\,6369. These
  values of luminosity are considerably lower than found in the recent
  literature. For NGC\,3242 Pauldrach et al.(\cite{paul}) find a value
  of 3200L\smallsun~while Kudritzki et al.(\cite{kud}) find
  10,000L\smallsun. For NGC\,6369 Monteiro et al.(\cite{mont}) find
  8100L\smallsun, but in this case much of the difference is due to
  the larger distance used by these authors.

\section{Discussion and Conclusions}

The nebular abundances of nine elements have been determined for the
PN NGC\,3242; abundances of the same elements, except for carbon, have
been found for NGC\,6369. Both nebulae are classified as having an
elliptical shape. The abundances found for both nebulae are very
similar. For helium, oxygen and neon they are essentially the same as
in the Sun.  This is consistent with the expectation that these PNe
are rather local. This is because the gradient in the stellar
abundances as a function of distance from the galactic center (e.g.
see Pottasch \& Bernard-Salas \cite{pbs}) would lead to the
expectation that a different abundance prevails when the position of
these PNe is non-local.

The state of evolution of these nebulae is considered in conjunction
with five other local PNe with elliptical shape. A comparison is made
between the observed abundances and those expected using the stellar
evolution models calculated by Karakas (\cite{karakas}). This
comparison leads to the conclusion that NGC\,3242 must be a descendent
of a low mass star, probably between 1M\smallsun~and 1.5M\smallsun.
NGC\,6369 probably has a slightly higher initial mass, but this conclusion
is less certain because the abundance of carbon is unknown.

The stellar temperature and luminosity is considered with the help of
the Zanstra method, the Energy Balance method and the observed nebular
excitation.  For NGC\,3242 we find T$_s$=80,000 K and
L=730L\smallsun~and for NGC\,6369 T$_s$=70,000 K and L=2400L\smallsun.
The temperatures are in quite good agreement with what is given in the
literature. The luminosities, which are distance dependent, are
considerably lower than given by the theories of Kudritzki et al.
(\cite{kud}) or Pauldrach et al. (\cite{paul}). This has already been
discussed by Napiwotzki (\cite{napi}) in a general way and by
Surendiranath \& Pottasch (\cite{nath}) for the case of NGC\,6826.
This is evidence that all of the nebulae listed in Table 15 have
central stars of quite low luminosity, probably between
500L\smallsun~and 2000L\smallsun.

The abundances of the other elements, S, Cl, P and Fe are all
lower than solar. For the first two of these elements the difference
with respect to the sun is a factor of two to three. In this case we
cannot distinguish between the possibility that this is caused by a
poor determination of the solar abundance or by the possibility that
important amounts of these elements have condensed in the form of
dust. For the remaining elements, P and Fe, the underabundance with
respect to the Sun is larger; it is likely that both of these elements
have formed molecules which have condensed in the form of dust. This
is probably true for all the PNe listed in Table 15.

\section{Acknowledgement}

This work is based on observations made with the Spitzer Space
Telescope, which is operated by the Jet Propulsion Laboratory,
California Institute of Technology under NASA contract 1407. Support
for this work was provided by NASA through Contract Number 1257184
issued by JPL/Caltech. We acknowledge that the IRS data on NGC\,6369 has been
taken from one of the GTO programs of Dale Cruikshank.


\begin{thebibliography}{}

\bibitem[1992]{acker}
Acker, A., Marcout, J., Ochsenbein, F. et al. 1992, Strasbourg-ESO catalogue
\bibitem[2003]{ackern}
Acker, A. \& Neiner, C. 2003, A\&A 403, 659
\bibitem[1979]{aller}
Aller, L.H., \& Czyzak, S.J. 1979, Ap\&SS 62, 397
\bibitem[1987]{allerkey}
Aller, L.H. \& Keyes, C.D. 1987, ApJS, 65,405
\bibitem[2005]{asplund}
Asplund, M., Grevesse, N., \& Sauval, A.J. 2005, ASP Conf.\,Ser. (Bash \& Barnes eds.)
\bibitem[1985]{barker}
Barker, T. 1985, ApJ 294, 193
\bibitem[1991]{becker}
Becker, R.H., White, R.L., \& Edwards, A.L. 1991, ApJS 75, 1
\bibitem[1999]{benjamin}
Benjamin, R.A., Skillman, E.D., \& Smits, D.P. 1999, ApJ 514, 307
\bibitem[2001]{bernard}
Bernard Salas, J., Pottasch, S.R., Beintema, D.A., \& Wesselius, P.R. 2001, A\&A 367, 949
\bibitem[2008]{bsp}
Bernard-Salas, J., Pottasch, S.R., Gutenkunst, S., Morris, P.W.,
Houck, J.R. 2008, ApJ, 672, 274
\bibitem[1992]{cahn}
Cahn, J.H., Kaler, J.B., \& Stanghellini, L. 1992, A\&AS 94, 399
\bibitem[1994]{fluks}
Fluks, M.A., Plez, B., de Winter, D., et al. 1994, A\&AS 105, 311
\bibitem[1988]{gath}
Gathier, R., \& Pottasch, S.R. 1988, A\&A 197, 266
\bibitem[1994]{grif}
Griffith, M.R., Wright, A.E., Burke, B.F. et al. 1994, ApJS 90, 179
\bibitem[1995]{hajian}
Hajian, A.R., Terzian, Y., \& Bignell, C. 1995, AJ 85, 2600
\bibitem[2000]{henry}
Henry, R.B.C., Kwitter, K.B., \& Bates, J.A. 2000, ApJ 531, 928
\bibitem[2004]{higdon}
Higdon, S.J.U., Devost, D., Higdon, J.L. et al. 2004, PASP 116, 975
\bibitem[2002]{hony}
Hony, S., Bouwman, J. Keller, L.P. et al. 2002, A\&A 393, L103
\bibitem[2004]{houck}
Houck, J.R., Appelton, P.N., Armus, L., et al. 2004, ApJS 154, 18
\bibitem[1987]{hummer}
Hummer, D.G., \& Storey, P.J. 1987, MNRAS 224, 801
\bibitem[2003]{karakas}
Karakas, A.I. 2003, Thesis, Monash Univ. Melbourne
\bibitem[2003]{kerber}
Kerber, F., Mignani, R.P., Guglielmetti, F. et al. 2003, A\&A 408, 1029
\bibitem[1994]{kings}
Kingsburgh, R.L., \& Barlow, M.J. 1994, MNRAS 271, 257
\bibitem[1997]{kud}
Kudritzki, R.-P., Mendez, R.H., Puls, J., \& McCarthy, J.K. 1997, IAU\,Symp.180, 64
\bibitem[2006]{krab}
Krabbe, A.C. \& Copetti, M.V.F. 2006. A\&A 450, 159
\bibitem[2001]{liulws}
Liu, X.-W., Barlow, M.J., Cohen, M. et al. 2001, MNRAS 323, 343
\bibitem[2004]{mell}
Mellema, G. 2004, A\&A 416, 623
\bibitem[1975]{milne}
Milne, D.K., \& Aller, L.H. 1975, A\&A 38, 183
\bibitem[2004]{mont}
Monteiro, H., Schwarz, H.E. Gruenwald, R. et al. 2004, ApJ 609, 194
\bibitem[2006]{napi}
Napiwotzki, R. 2006, A\&A 451, L27
\bibitem[2004]{paul}
Pauldrach, A.W.A., Hoffmann, T.L., \& Mendez, R.H. 2004, A\&A 419, 1111 
\bibitem[2001]{pena}
Pena, M., Stasinska, G., \& Medina, S. 2001, A\&A 367, 983
\bibitem[2003]{phillips}
Phillips, J.P. 2003, MNRAS 344, 501
\bibitem[2005]{porter}
Porter, R.L., Bauman, R.P., Ferland, G.J., et al. 2005, ApJ 622, L73
\bibitem[1984]{potta}
Pottasch, S.R. 1984, Planetary Nebulae, Reidel Publ. Co. (Dordrecht)
\bibitem[1999]{pott1}
Pottasch, S.R., \& Beintema, D.A. 1999, A\&A 347, 974
\bibitem[2000]{pott2}
Pottasch, S.R., Beintema, D.A., \& Feibelman, W.A. 2000, A\&A 363, 767
\bibitem[2001]{pott4}
Pottasch, S.R., Beintema, D.A., Bernard Salas, J., \& Feibelman,
W.A. 2001, A\&A 380, 684
\bibitem[2004]{pott6}
Pottasch, S.R., Bernard-Salas, J., Beintema, D.A. \& Feibelman, W.A. 2004, A\&A
423, 593
\bibitem[2005]{pott7}
Pottasch, S.R., Beintema, D.A. \& Feibelman, W.A. 2005, A\&A 436, 965
\bibitem[2006]{pbs}
Pottasch, S.R., \& Bernard-Salas, J. 2006, A\&A 457, 189
\bibitem[2008]{pott8}
Pottasch, S.R., Bernard-Salas, J., Roellig, T.L. 2008, A\&A 481, 393
\bibitem[1983]{pmp}
Preite-Martinez, A., \& Pottasch, S.R. 1983 A\&A, 126, 31
\bibitem[1979]{seaton}
Seaton, M.J. 1979, MNRAS 187, 73
\bibitem[2008]{nath}
Surendiranath, R., \& Pottasch, S.R. 2008, A\&A 483, 519
\bibitem[1997]{terz}
Terzian, Y. 1997, IAU Symp.180, p.29 (eds. H.J.Habing \& H.Lamers)
\bibitem[2002]{tinkler}
Tinkler, C.M. \& Lamers, H.J.G.L.M. 2002, A\&A 384, 987
\bibitem[2003]{tsamis}
Tsamis, Y.G., Barlow, M.J., Liu. X.-W., et al. 2003, MNRAS 345, 181
\bibitem[2004]{werner}
Werner, M., Roellig, T.L., Low, F.J., et al. 2004, ApJS 154, 1


\end{thebibliography}
\end{document}